\documentclass[review]{elsarticle}

\usepackage{lineno,hyperref}
\usepackage{amsmath}
\usepackage{amssymb}
\usepackage{yhmath}
\usepackage{booktabs, multirow} 
\usepackage{graphicx,psfrag,epsf}
\usepackage{float}
\usepackage{caption}
\usepackage{subcaption} 
\graphicspath{ {./figures/} }
\usepackage{xcolor} %text color
\usepackage{hyperref}
\hypersetup{colorlinks,
citecolor=blue
}
\usepackage{geometry}
\makeatletter
\def\ps@pprintTitle{%
  \let\@oddhead\@empty
  \let\@evenhead\@empty
  \let\@oddfoot\@empty
  \let\@evenfoot\@oddfoot
}
\makeatother

\modulolinenumbers[5]

\begin{document}
\include{macros} % include my custom macros

\begin{frontmatter}

\title{A Multivariate Point Process Model for\\ Simultaneously Recorded Neural Spike Trains}

\author[mainaddress]{Reza Ramezan\corref{mycorrespondingauthor}}
\cortext[mycorrespondingauthor]{Corresponding author}
\ead{rramezan@uwaterloo.ca}

\author[mainaddress]{Meixi Chen}
\author[mainaddress]{Martin Lysy}
\author[mainaddress]{Paul Marriott}

\address[mainaddress]{Department of Statistics and Actuarial Science, University of Waterloo, Canada}

\begin{abstract}
The current state-of-the-art in neurophysiological data collection allows for simultaneous recording of tens to hundreds of neurons, for which point processes are an appropriate statistical modelling framework. However, existing point process models lack multivariate generalizations which are both flexible and computationally tractable. This paper introduces a multivariate generalization of the Skellam process with resetting (SPR), a point process tailored to model individual neural spike trains. The multivariate SPR (MSPR) is biologically justified as it mimics the process of neural integration. Its flexible dependence structure and a fast parameter estimation method make it well-suited for the analysis of simultaneously recorded spike trains from multiple neurons. The strengths and weaknesses of the MSPR are demonstrated through simulation and analysis of experimental data.
\medskip
\end{abstract}


\end{frontmatter}

\section{Introduction}

The statistical modelling of neural spike trains falls under the general framework of point processes. Among point processes, the (inhomogeneous) Poisson process is a popular choice \cite{Brown1998,Ramezan2014}, but its theoretical properties 
% (unless modifications are made) 
fail to address biological phenomena such as refractoriness and burst spiking activities \cite{Ramezan2016}. Furthermore, the Poisson process has very limited multivariate generalizations in terms of dependence structure \cite{Marshall1985, Kocherlakota1992, Johnson1997, Lakshminarayana1999, Karlis2003, Karlis2005}. Some alternative models include Integrate-and-Fire models \cite{Gerstein1964, Paninski2008}, GLMs \cite{Kass2001, Dimatteo2001,Truccolo2005, Wojcik2009}, renewal processes \cite{Koyama2014}, and latent variable models \cite{Shahbaba2014, Zhou2016}.\

The Skellam process with resetting (SPR) is a point process recently developed to model spike trains collected from an individual neuron \cite{Ramezan2016}. A Skellam process $\{S(t), t\geq0\}$ is defined as the difference between two independent Poisson processes, i.e. $S(t)=N_1(t)-N_2(t)$, where  $N_i(t)\sim Poisson(\lambda_i\,t)$, $i=1,2$.  Let $T_n = \min\{t: S(t) \ge n\}$. The SPR processs is then defined as $\tilde S(t) = \{S(t) - \sum_{j=1}^\infty \delta(t \ge T_j),~t\geq0\}$, where $\delta(\cdot)$ is the indicator function. Thus, resetting refers to bringing the path of the Skellam process back to state 0 after each spike, i.e. after each record of the process, to mimic the refractoriness of neurons \cite{Ramezan2016}.  In this paper we generalize the results of \cite{Ramezan2016} and introduce the multivariate Skellam process with resetting (MSPR) for the analysis of simultaneously recorded spike trains from multiple neurons. 

\section{Multivariate Skellam Process with Resetting}

We first introduce the multivariate Skellam random vector, following a similar construction to that the multivariate Poisson random vector \cite{Karlis2003, Karlis2005}.\

{\bf Definition 1:} Let $\mathbf{Z}$ be a $p$-variate random vector where
\begin{equation}
\mathbf{Z} =    \begin{pmatrix}
                Z_1\\
                Z_2\\
                \vdots \\
                Z_i \\
                \vdots \\
                Z_p \\
                \end{pmatrix}=
                \begin{pmatrix}
				Y_1 &+& \sum_{j=2}^p a_{1j} Y_{1j} && \\
				Y_2 &+& a_{21}Y_{12} &+& \sum_{j=3}^p a_{2j} Y_{2j} \\ 
				\vdots \\
				Y_i &+& \sum_{j=1}^{i-1} a_{ij} Y_{ji} &+& \sum_{j=i+1}^p a_{ij} Y_{ij} \\
				\vdots \\
				Y_p && &+& \sum_{j=1}^{p-1} a_{pj} Y_{jp}\\
				\end{pmatrix},\nonumber
\end{equation}
where $a_{ij}\in\{-1,0,1\}$, $Y_{i}\sim Skellam(\lambda_{i1},\lambda_{i2})$, and  $Y_{ij}\sim Skellam(\gamma_{ij},\gamma_{ij})$. All Skellam random variables $Y_i$ and $Y_{ij}$, $\forall i,j$ are independent of one another, and also $a_{ij}=0\implies a_{ji}=0$. The random vector $\mathbf{Z} \sim MSk(\boldsymbol{\lambda}_1, \boldsymbol{\lambda}_2, \boldsymbol{\gamma})$ is called a {\em multivariate Skellam random vector} with parameters $\boldsymbol\lambda_{1}=(\lambda_{11},...,\lambda_{p1})$, $\boldsymbol\lambda_{2}=(\lambda_{12},...,\lambda_{p2})$, and $\boldsymbol\gamma=(\gamma_{12},...,\gamma_{(p-1)p})$.\

{\bf Definition 2:} Let $\mathbf{S}(t) = \{S^{(i)}(t),t\geq0, i=1,\ldots,p\}$ be a $p$-dimensional Skellam processes, i.e., using Definition 1 with Skellam processes instead of random variables. Then for $T^{(i)}_{n} = \min\{t : S^{(i)}(t) \ge n\}$, the {\em multivariate Skellam process with resetting (MSPR)} is defined as
$$
\tilde{\mathbf{S}}(t) =\Big\{S_0^{(i)}(t)-\sum_{j=1}^\infty \delta(t \ge   T^{(i)}_{j} ),~t\geq0,~i=1,\ldots,p\Big\}.
$$
The value of a $p$-variate Skellam process at time $t=t_0$ is a multivariate Skellam random vector, $\mathbf{S}(t_0)\sim MSk(\boldsymbol\lambda_1t_0,\boldsymbol\lambda_2t_0,\boldsymbol\gamma\,t_0)$. Within this process, the spike times of neuron $i$ are modelled as the record times of the $i$th SPR within the multivariate Skellam process. These individual processes are potentially dependent to allow modelling the relationships between cells in a neuronal ensemble. 

To the best of our knowledge, the MSPR is the first multivariate point process model for neural spike trains which allows for the full range of correlation coefficient. Indeed, neurons $i$ and $j$ are positively correlated if $sign(a_{ij})=sign(a_{ji})$, negatively correlated if $sign(a_{ij})=-sign(a_{ji})$, and uncorrelated if $a_{ij}=a_{ji}=0$. 

\subsection{Parameter Estimation}
We assume that the spike times from multiple independent trials are available for each of the simultaneously recorded neurons. The duration of each trial is assumed to be ``long'' relative to the average inter-spike interval so that asymptotic results apply.\

Following Definitions 1 and 2, and the parameterization of \cite{Ramezan2016}, the two parameters of the marginal SPR for neuron $i$ are  $\lambda_1^{(i)}=\lambda_{i1}+\sum_{j}\gamma_{ij}$, $\lambda_2^{(i)}=\lambda_{i2}+\sum_{j}\gamma_{ij}$, for $i=1,..,p$ .  We use the method of moments to estimate $\gamma_{ij}$ using spike counts in each trial, i.e. $\widehat\gamma_{ij}=|0.5\widehat\sigma_{ij}|$ in which $\widehat\sigma_{ij}$ is the estimate of the covariance between spike counts of the two neurons $i$ and $j$. We then plug in the moment estimate $\widehat\gamma_{ij}$ in the likelihood function of the marginal SPR processes and compute maximum profile likelihood estimates of parameters $\lambda_1^{(i)}$ and $\lambda_2^{(i)}$ for $i=1,...,p$ using constrained optimization.  The standard error of estimates are obtained by boostrapping over the independent trials.

\section{Data Analysis}
We assessed the MSPR model using a 3-neuron simulation study as well as a 5-neuron experimental dataset from the prefrontal cortex of a mouse (3 in OFC and 2 in PL) in a classical conditioning study where the observation window is 10 seconds after a reward delivery \cite{Kaminska2020}.

\begin{table}[!htb]
\begin{center} 

\caption{Parameter estimation for 50 repeated trials from an ensemble of three simulated neurons} \label{Simulation-Estimation} 

{\small
\begin{tabular}{llllllllll} 
%\toprule
\hline
& $\lambda_{11}$ & $\lambda_{12}$ & $\lambda_{21}$ & $\lambda_{22}$ & $\lambda_{31}$ & $\lambda_{32}$ & $\gamma_{12}$ & $\gamma_{13}$ & $\gamma_{23}$\\
%\midrule
\cmidrule{2-10}
True & 15 & 10 & 20 & 15 & 10 & 7 & 5 & 15 & 10\\
Est. & 17.8 & 11.6 & 19.7 & 12.9 & 15.2 & 10.9 & 5.0 & 11.5 & 8.3 \\
SE & 7.4 & 7.2 & 6.9 & 7.0 & 7.8 & 7.6 & 3.2 & 5.2 & 5.7 \\
%\bottomrule
\hline
\end{tabular} 
}
\end{center} 
\end{table}

\begin{table}[!ht]
\begin{center} 

\caption{Moment estimates of the ISIs from 13 repeated trials from an ensemble of five prefrontal cortex neurons} \label{Real-Estimation} 

{\small
\begin{tabular}{llll} 
%\toprule
\hline
 && \multicolumn{1}{c}{Mean} & \multicolumn{1}{c}{Variance}   \\
\cmidrule{2-3} \cmidrule{4-4}
 & Observed ISI (SE) & 0.075 (0.035) & 0.007 (0.008)   \\
 \multirow{-2}{*}{Neuron \#1 (OFC)}
& Model ISI (SE) & 0.121 (0.069) & 0.015 (0.025)   \\
 & Observed ISI (SE) & 0.059 (0.011) & 0.003 (0.002)   \\
 \multirow{-2}{*}{Neuron \#2 (OFC)}
& Model ISI (SE) & 0.096 (0.021) & 0.009 (0.004)   \\
 & Observed ISI (SE) & 0.054 (0.011) & 0.002 (0.001)   \\
 \multirow{-2}{*}{Neuron \#3 (OFC)}
& Model ISI (SE) & 0.083 (0.021) & 0.007 (0.004)   \\
 & Observed ISI (SE) & 0.076 (0.025) & 0.009 (0.008)   \\
 \multirow{-2}{*}{Neuron \#4 (PL)}
& Model ISI (SE) & 0.118 (0.047) & 0.014 (0.014)   \\
 & Observed ISI (SE) & 0.057 (0.021) & 0.007 (0.006)   \\
 \multirow{-2}{*}{Neuron \#5 (PL)}
& Model ISI (SE) & 0.080 (0.037) & 0.006 (0.008)   \\
%\bottomrule
\hline
\end{tabular} 
}
\end{center} 
\vspace{-.2cm}
\end{table}

\begin{figure}[ht!]
\begin{center}
\includegraphics[scale=0.4]{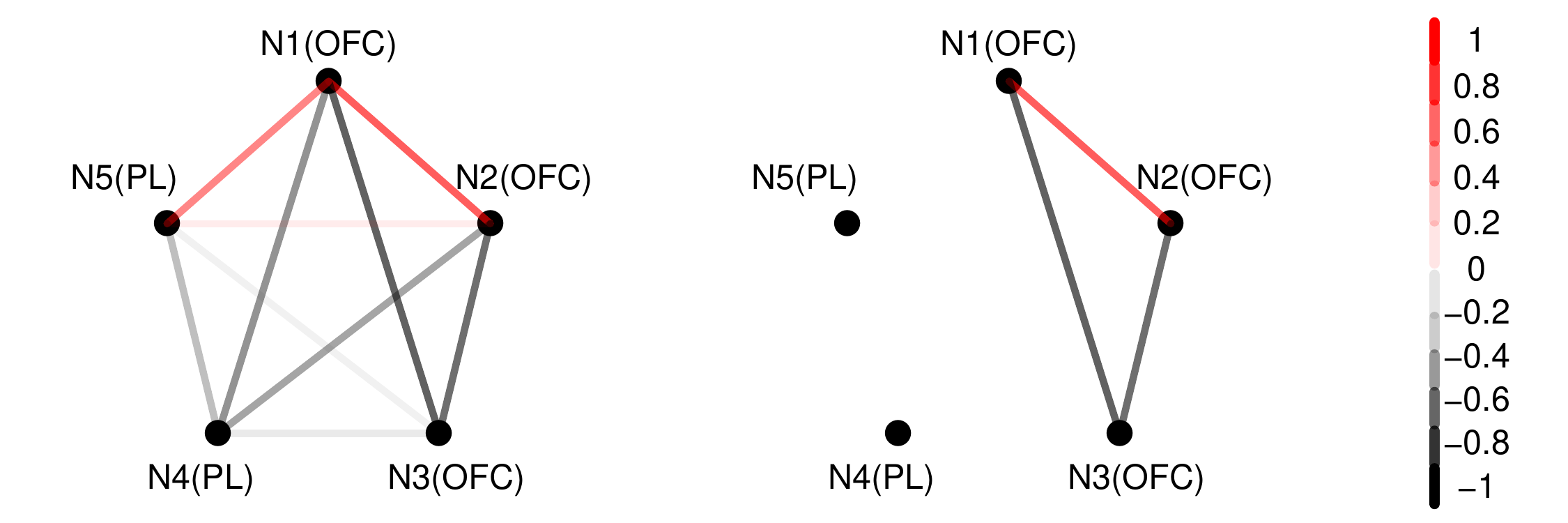}
\end{center}
\caption{
Spike count correlation (left) and statistically significant correlations only (right).
} 
\label{Real-figure}
\end{figure}

 Table~\ref{Simulation-Estimation} shows the results for 50 spike trains from a 3-neuron simulation study. Table \ref{Real-Estimation} and Figure \ref{Real-figure} present the results for the 5-neuron experiment.  To put the parameter estimates into context, Table \ref{Real-Estimation} displays the means and variances of the inter-spike intervals (ISI) as well as those estimated by the MSPR model.  Comparing the mean and variance estimates from the model to those of the empirical ISI data, it seems that a slight systematic upward bias exists in the model fit. While PP-plots for each neuron (not shown here) did not reveal a noticeable lack of fit, bias correction for the parameter estimation method could potentially be explored.

 Figure \ref{Real-figure} suggests that all OFC neurons are correlated, whereas PL neurons are not. The correlations between OFC and PL neurons were not statistically significant.

\section{Discussion}

This paper introduces the multivariate Skellam process with resetting for the analysis of simultaneously recorded neural spike trains. Unlike other multivariate point processes, the MSPR has a flexible dependency structure and can be fit with relative ease. Future work includes the additional modelling of the crossing times $T_n^{(i)} = \min\{t: S^{(i)} \ge nk_i\}$ for $k_i$, $i=1,\ldots,p$ to be estimated from the data, as was done in the univariate SPR model of \cite{Ramezan2016}.  The challenge is to perform parameter estimation under this more complicated model without sacrificing computational efficiency.

\section{Acknowledgments}
 This work was supported by the Natural Sciences and Engineering Research Council of Canada. We thank David Moorman of UMass Amherst for the classical conditioning data.

%\section*{References}

\bibliographystyle{plain}
%\spacingset{1}
\bibliography{RamezanEtAl2022}
	
\end{document}